\documentclass[prl,twocolumn,groupaddress,showpacs]{revtex4}
\usepackage{epsfig,amsmath,amssymb,graphics,color,calc}
\newcommand{\be}{\begin{equation}}
\newcommand{\ee}{\end{equation}}
\renewcommand{\phi}{\varphi}
\newcommand{\ba}{\begin{eqnarray}}
\newcommand{\ea}{\end{eqnarray}}

\begin{document} 
\title{Influence of the glass transition on the liquid-gas spinodal
decomposition}

\author{Vincent Testard}
\affiliation{Laboratoire Charles Coulomb, 
UMR 5221 CNRS and Universit\'e Montpellier 2, Montpellier, France}

\author{Ludovic Berthier}
\affiliation{Laboratoire Charles Coulomb, 
UMR 5221 CNRS and Universit\'e Montpellier 2, Montpellier, France}

\author{Walter Kob}
\affiliation{Laboratoire Charles Coulomb, 
UMR 5221 CNRS and Universit\'e Montpellier 2, Montpellier, France}

\date{\today}
\begin{abstract}
We use large-scale molecular dynamics simulations 
to study the kinetics of the liquid-gas phase separation if the
temperature is lowered across the glass transition 
of the dense phase. We observe a gradual change from 
phase separated systems at high temperatures to nonequilibrium,
gel-like structures that evolve very slowly at low temperatures. 
The microscopic mechanisms responsible for the coarsening 
strongly depend on temperature, and change from 
diffusive motion at high temperature to a strongly
intermittent, heterogeneous and thermally activated dynamics 
at low temperature, leading to logarithmically slow growth 
of the typical domain size. 
\end{abstract}



\maketitle

The kinetics of phase separation is a fascinating 
subject with a long history in condensed matter physics~\cite{hohenberg}. 
Although this nonequilibrium phenomenon is 
complex, universal scaling behaviour 
have been indentified and can be studied using statistical 
mechanics~\cite{bray}. Phase separation kinetics also has 
important experimental and technological consequences~\cite{tanaka}. 
While this phenomenon is well-understood for simple materials, 
more complex behaviour can be expected if the different
phases have specific properties, such as viscolelastic 
polymer solutions or colloidal suspensions, 
which interfere and possibly modify the coarsening process
in a non-trivial manner. 

Here we study numerically the kinetics 
of the liquid-gas spinodal decomposition over a 
broad temperature range encompassing the glass transition 
of the dense fluid. Not much is known about the resulting gas-glass spinodal
decomposition--a specific instance of a viscoelastic
phase separation~\cite{tanaka}--,
although this interplay is invoked in a number 
of contexts~\cite{tanaka,danchinov}. Further interesting cases are
colloids~\cite{lu} and proteins~\cite{cardinaux}
with short-ranged attraction that form a gel, instead of phase separating. 
A plausible interpretation is that
the dense phase is actually a glass, and this `arrests'
the phase separation~\cite{lu,cardinaux}. 
It has recently been suggested that similar structures could be obtained 
in C$_{60}$ molecules~\cite{paddyc60}, and that the range of the 
potential is important~\cite{paul}.
Bicontinous disordered structures
reminiscent of the ones obtained in spinodal decompositions
may also be found in colourful 
bird feathers, and were recently interpreted 
as incompletely phase separated polymeric 
glasses~\cite{dufresne}. 
Although it can be expected that the glass transition 
slows down the spinodal decomposition, the general 
mechanisms at play have never been studied in any detail
on the microscopic scale.

Although coarsening processes are most efficiently studied theoretically
using coarse-grained models, such as model H for the 
liquid-gas spinodal decomposition~\cite{hohenberg}, in practice
it is difficult to faithfully incorporate
the complex---typically highly nonlinear and history 
dependent---physical properties of glasses in such 
descriptions~\cite{jackle}.
Thus, we work in the opposite direction, and start from 
a microscopically realistic description of the homogeneous glass
and study its behaviour during phase separation.
Indeed, numerous successful atomic-scale simulations of 
the liquid-gas spinodal decomposition 
exist~\cite{yamamoto,koch,velasco,laradji}. 
In Ref.~\cite{evans}, a Lennard-Jones system was quenched 
to low temperature in the coexistence region, and the resulting  
gas-crystal phase separation was studied, 
but no arrest was reported. 
Simulations of realistic colloidal interactions were 
reported~\cite{foffi,dave,zacca},
but the quenches have been performed at very low temperatures 
where particle aggregation are nearly irreversible 
and thermal fluctuations play little role. 

We use molecular dynamics simulations to 
study a 80:20 binary Lennard-Jones mixture
with interaction parameters chosen to yield excellent 
glass-forming ability~\cite{KA}.
Specific attention was paid here 
to system sizes. While numerical studies
of the glass transition in the homogeneous liquid 
typically require simulating about $10^3$ particles, 
we found that up to $10^6$ particles were needed to obtain 
results devoid of finite size effects during the phase
separation. We obtained most of our quantitative 
results using $3 \cdot 10^5$ particles. 
We study the kinetics of the phase separation 
following an instantaneous quench at constant density from
the high temperature fluid phase to the coexistence region, 
which we determined for the present system. In the following
we use Lennard-Jones units corresponding to the majority
component, expressing length in units of particle diameter, $\sigma$, 
and time in units of $\tau = \sqrt{m \sigma^2 / \epsilon}$,
where $m$ is the particle mass, and $\epsilon$ the energy scale
in the Lennard-Jones interaction.

\begin{figure}
\psfig{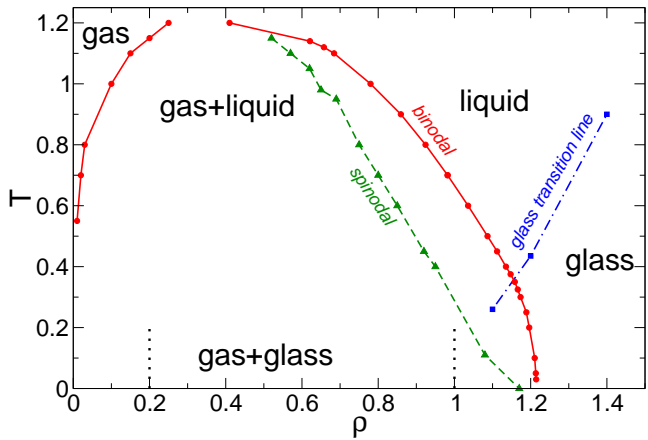}
\caption{\label{diag} Phase diagram of the binary Lennard-Jones mixture
showing the location of the homogeneous gas, liquid, and glass phases,
and the gas-liquid and gas-glass coexistence regions. The glass line 
corresponds to the mode-coupling temperature (Ref.~\cite{gillesmct}), 
while the spinodal
is taken from Ref.~\cite{sastry}. The vertical dashed lines separate 
disconnected droplets from bicontinuous structures.}
\end{figure}

In Fig.~\ref{diag} we present a density, $\rho$, and 
temperature, $T$, phase diagram showing the location of the 
homogeneous gas, liquid, and glass phases of the system, 
as well as the coexistence region. For the glass `transition',
we show the location of the temperature $T_{\rm mct}$
obtained numerically~\cite{gillesmct} from fitting the temperature evolution   
of the relaxation time of the system using predictions from 
the mode-coupling theory. Usually, 
it is hard to maintain thermal 
equilibrium below $T_{\rm mct}$, which can be taken 
as a proxy for the `computer' glass temperature. 
Figure~\ref{diag} demonstrates that the present model exhibits 
the right ingredients, as in 
a quench below $T \approx 0.3$ in the coexistence region, 
the dense phase is a glass---not a liquid.

On the liquid side, the coexistence line
hits the glass line near $(\rho \approx 1.15, T \approx 0.3)$. 
The glass line can be followed between  the binodal and the spinodal, 
where the homogeneous liquid is metastable. Using methods described
below, we found that the coexistence
line is only weakly affected by the glass line, as found 
in some experiment~\cite{lu}. We do not
find a reentrant coexistence line slaved 
to the glass line, as reported in Ref.~\cite{cardinaux}. 

\begin{figure}
\psfig{file=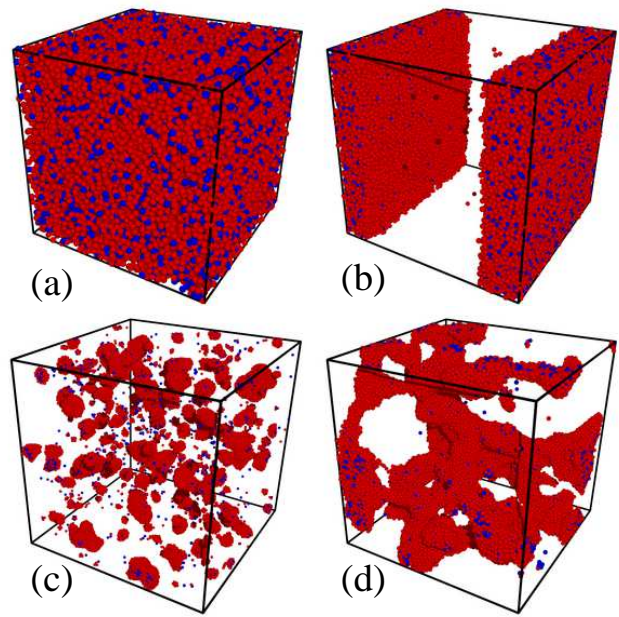,width=8.5cm,clip}
\caption{\label{image} Representative structures of 
(a) the liquid phase at $(T=1.75, \rho=0.6)$, (b) a fully demixed system
at $(T=0.5, \rho=0.4)$, (c) disconnected growing droplets
at $(T=0.1, \rho=0.1, t=10^4)$ (d) a bicontinous gel-like structure 
obtained at long time at $(T=0.1, \rho=0.2, t=10^4)$ where phase 
separation is incomplete.}
\end{figure}

In Fig.~\ref{image} we present typical configurations 
obtained in various parts of the phase diagram. 
Figure~\ref{image}a shows 
a homogeneous configuration obtained in the liquid phase, while 
Fig.~\ref{image}b shows the final configuration obtained after 
a quench to $(T=0.5, \rho=0.4)$, where the system is fully demixed
at long times. It is in this temperature regime that 
the liquid-gas spinodal decomposition is 
typically studied~\cite{koch,yamamoto,laradji,velasco}. 
For a quench to densities below $\rho \approx 0.2$ we find that
the coarsening proceeds via growth of disconnected 
droplets, as illustrated in Fig.~\ref{image}c. 
Similarly, we find that for densities above  
$\rho \approx 1.0$, the coarsening proceeds via the
growth of disconnected gas bubbles immersed in a dense fluid.
More interesting is the low temperature regime in the density range
$\rho \in [0.2, 0.8]$ where bicontinuous, percolating, gel-like structures 
are obtained at long times, see Fig.~\ref{image}d. This means 
that phase separation is 
incomplete when $T$ is low, at least in the timescale
of the simulation. These 
configurations are strikingly reminiscent of the 
confocal microscopy images in colloidal gels~\cite{lu,paul}. 

To study quantitatively the
phase separation kinetics, we must characterize the structures 
shown in Fig.~\ref{image}d and follow their time dependence 
for a quench at a given state point.
Phase separating systems are usually characterized 
in numerical work using the static structure factor
of density fluctuations, or equivalently, the pair correlation
function. However, we found that a
numerically more efficient characterization of the bicontinuous
structures is provided by the chord length distribution
of the low-density phase, which contains detailed 
information about the domain size distribution. This is 
a standard characterization tool for 
porous media~\cite{levitz,gubbins}.

To measure this quantity, we must first numerically determine the location
of the interfaces separating the two phases. This requires a
non-trivial coarse-graining for a particle-based simulation 
in which complex topologies are present. 
Thus, we discretize space in cubes of small linear 
size $d=0.5\sigma$, and define on this cubic lattice
a coarse-grained density field, $\bar\rho({\bf r})$, 
using a weighted average over cells around ${\bf r}$. 
The distribution of the coarse-grained 
density, $p(\bar \rho)$, is typically bimodal, as expected 
for a two-phase system. In particular, 
$p(\bar \rho)$ exhibits at large $\bar \rho$ a maximum, which 
provides an accurate measure of the average density
of the dense phase. We have used this maximum to determine 
the coexistence line at low temperatures in the phase 
diagram of Fig.~\ref{diag}.  Moreover,
a careful analysis of $p(\bar \rho)$, backed by direct visualizations, 
shows that a density threshold can be estimated to 
accurately locate gas and fluid domains, 
and the position of the interfaces separating them. 

\begin{figure}
\psfig{file=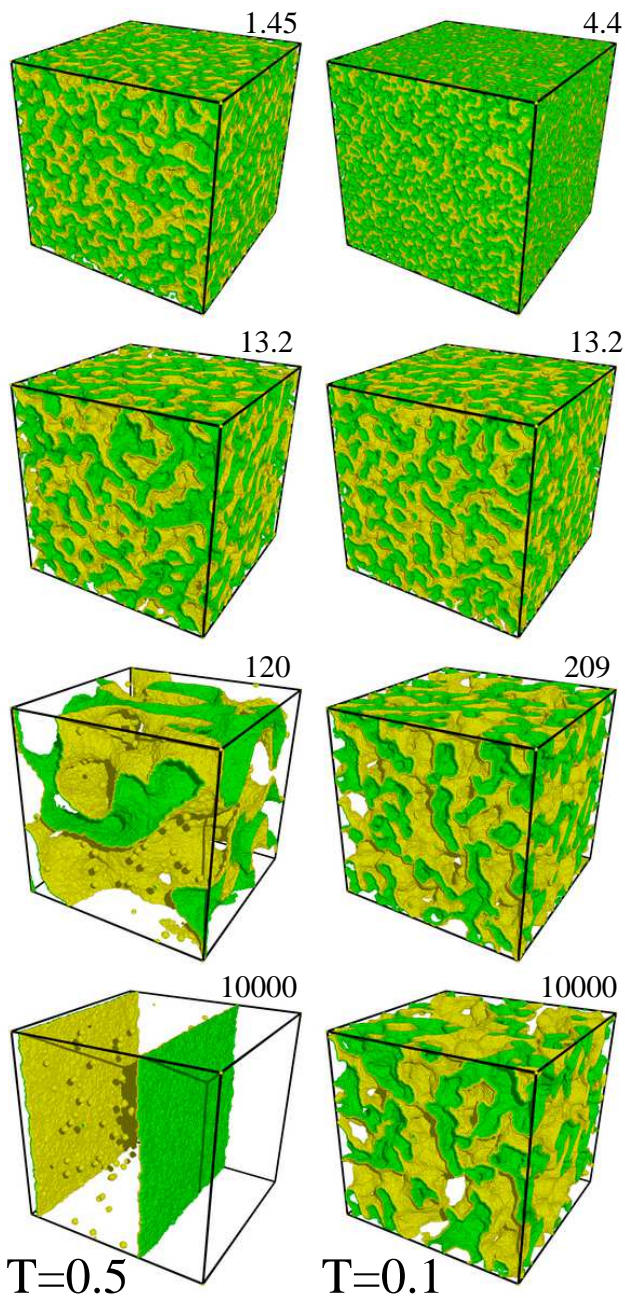,width=8.5cm,clip}
\caption{\label{spinodal} 
Time series for a quench at $\rho=0.4$ showing 
interfaces, the yellow/green side 
being inside the gas/fluid phase.
While the liquid-gas spinodal decomposition
at $T=0.5$ proceeds until complete phase separation, dynamics
is strongly slowed down in the gas-glass case at 
$T=0.1$.}
\end{figure}

The time evolution of the so obtained 
interfaces is shown in Fig.~\ref{spinodal} after a quench 
at $\rho=0.4$ in the coexistence region.
At $T=0.5$, the spinodal decomposition proceeds 
rapidly until complete phase separation of the system.
By contrast, for $T=0.1$ the evolution is similar to the one at $T=0.5$ at 
short times, $t < 10^2$, but further evolution
is strongly suppressed (but not fully arrested) 
over the remaining two decades of the simulations, up to $t=10^4$. 

We are now in a position to measure the chord length distribution
and its evolution. Chords are defined by two consecutives intersections
of a straight line with the interfaces of the two-phase material. 
In practice, we measure chords along the three axis of the lattice  
used to coarse-grain the density, and measure the length
$\ell$ of the segments belonging to the gas phase. 
Thus, we obtain for each configuration a 
distribution $P(\ell,t)$. We find that 
$P(\ell,t)$ has a maximum, and beyond the maximum
is well-described by an exponential tail, as
commonly found in porous media~\cite{gubbins,levitz}. 
We define the average domain size as
the first moment of the distribution: 
$L(t) = \int_0^\infty \ell  P(\ell, t) d \ell$.
We find that $L(t)$ is a robust and efficient measure 
of the typical domain size, which
could be easily implemented in confocal microscopy 
experiments~\cite{lu,paul}.

\begin{figure}
\psfig{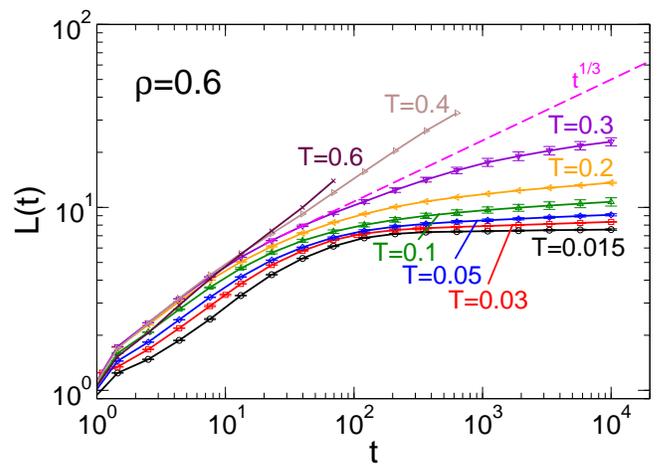}
\caption{\label{length} Evolution of the typical domain size
at $\rho=0.6$ and various $T$, in a log-log plot. 
Below $T \approx  0.3$, the growth is much slower than the diffusive
law, $L(t) \sim t^{\frac{1}{3}}$ (dashed line), but not fully arrested.}  
\end{figure}

In Fig.~\ref{length} we report the time evolution of $L(t)$ 
at density $\rho = 0.6$ and various temperatures from $T=0.6$ down
to very low temperatures, $T=0.015$. 
When temperature is large, we find that the growth law is well described 
by a power-law, $L(t) \sim t^n$, with an apparent 
exponent $n \approx \frac{1}{2}$~\cite{koch,yamamoto,velasco}. 
We interpret this behaviour
as the indication that our data are taken in the crossover
between the diffusive ($n = \frac{1}{3}$) 
and hydrodynamic ($n =1$) regimes, which are notoriously difficult
to disentangle in molecular dynamics simulations~\cite{laradji,lebo}. 

The most important observation in Fig.~\ref{length} is that 
for $T < 0.3$ the growth law is always much slower than the diffusive
$t^{1/3}$ law and strongly depends on temperature, 
as reported in a recent colloidal experiment~\cite{paul}.
The data are actually curved in a log-log 
representation, suggesting that the growth is not algebraic, but 
presumably logarithmic. We
emphasize that these data are free of finite size effects. 
Indeed we found that smaller system sizes 
yield asymptotic lengthscales that are smaller, and spinodal decomposition 
appears to be much more strongly slowed down 
at low $T$ if $N$ is chosen too small, a fact which 
is experimentally relevant for gelation in 
confinement~\cite{confined}. 
A closer look at the data at large times and low $T$ in 
Fig.~\ref{length} shows that if the growth is much slower 
than diffusive, it is not completely suppressed either, that is, 
spinodal decomposition is not kinetically arrested 
by the glass transition of the dense phase, as assumed in 
previous work~\cite{lu,cardinaux,foffi,paddyc60}, 
although it does become difficult to detect at very low temperatures. 

\begin{figure}
\psfig{file=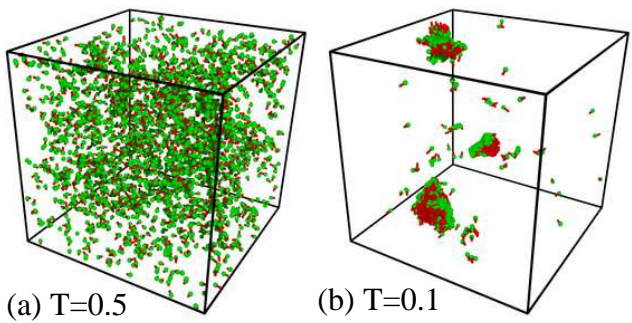,width=8.75cm,clip}
\caption{\label{fast} Mobile particles (green spheres) and their displacements 
(red cylinders) at $\rho=0.4$
for time intervals corresponding to a equivalent 
domain sizes. (a) $T=0.5$,  times $t=32$ and $t=34.5$. 
(b) $T=0.1$, times $t=1096$ and $t=1905$.}   
\end{figure}

Instead, we find that low temperatures, $T < 0.3$, 
strongly modify the microscopic
mechanisms of the phase separation.
While for the liquid-gas phase separation 
particles diffuse to relax the curvature of the interfaces, 
at low $T$ surface tension seems to play very little 
role. Direct visualization shows that curved interfaces
quickly solidify and hardly relax, while  
particle diffusion in the dense phase (an aging glass) is 
strongly suppressed. 

Particle displacements help to understand
the microscopic mechanisms at play. In a given
time interval, we select 
about 1~\% of the particles that have moved farthest. 
At high $T$
we find displacements as in Fig.~\ref{fast}a where 
mobile particles appear to be randomly scattered throughout the system, 
with uncorrelated displacements. The picture is very different at 
low $T$, see Fig.~\ref{fast}b, which 
reveals strong clustering of the fast moving particles, with 
very correlated displacements. 
At low temperature, particles in the 
dense glassy domains are basically arrested 
and can only move significantly if the whole domain 
moves as a rigid body. Such large 
displacements occur in a temporally intermittent
and spatially heterogeneous manner, and follow a change 
in the local topology of the bicontinuous structure, for instance 
when a thin neck breaks. There are three such relaxation events
in the example of Fig.~\ref{fast}b. We speculate that the main 
driving force of the phase separation process is the 
mechanical stress stored in these non-equilibrium, disordered, 
porous structures, rather than surface 
tension. Due to thermal fluctuations, these constraints are 
released in a highly heterogeneous manner, yielding large 
scale rearrangements with correlated particle displacements, 
which ultimately increase the average domain size.

In future work, one should characterize in more detail the 
microscopic aging dynamics of the system and extend our studies to
different kinds of models, as spinodal decomposition
represents a promising tool to produce disordered porous
media, even in atomic systems.  

We thank D. Reichman for initially stimulating this work, 
B. Coasne, T. Gibaud, and P. Royall for useful discussions.
This work is partially funded by ANR Dynhet and R\'egion 
Languedoc Roussillon.
W. K. is a member of Institut Universitaire de France.

\end{document}